%% file: main.tex
\documentclass[sigconf]{acmart}

\usepackage{amsmath}
\DeclareMathOperator*{\argmax}{arg\,max}

\usepackage{graphicx}
\usepackage{subfig}

\AtBeginDocument{%
  }

\copyrightyear{2024}
\acmYear{2024}
\setcopyright{rightsretained}
\acmConference[UbiComp Companion '24]{Companion of the 2024 ACM
International Joint Conference on Pervasive and Ubiquitous Computing }{October
5--9, 2024}{Melbourne, VIC, Australia}
\acmBooktitle{Companion of the 2024 ACM International Joint Conference on
Pervasive and Ubiquitous Computing (UbiComp Companion '24), October 5--9,
2024, Melbourne, VIC, Australia}
\acmDOI{10.1145/3675094.3679000}
\acmISBN{979-8-4007-1058-2/24/10}

\usepackage{algorithm}
\usepackage{algorithmic}
\usepackage{amsmath}
\usepackage{listings}

\begin{document}

\title[Using LLMs to Compare XAI methods for HAR in Smart Homes]{Using Large Language Models to Compare Explainable Models\\ for Smart Home Human Activity Recognition}

\author{Michele Fiori, Gabriele Civitarese, Claudio Bettini}
\email{{michele.fiori,gabriele.civitarese,claudio.bettini}@unimi.it}
\affiliation{%
  \institution{EveryWare Lab, Dept. of Computer Science, University of Milan}
  \city{Milan}
  \country{Italy}
}

\renewcommand{\shortauthors}{Fiori et al.}

\begin{abstract}
Recognizing daily activities with unobtrusive sensors in smart environments enables various healthcare applications. Monitoring how subjects perform activities at home and their changes over time can reveal early symptoms of health issues, such as cognitive decline.
Most approaches in this field use deep learning models, which are often seen as black boxes mapping sensor data to activities. However, non-expert users like clinicians need to trust and understand these models' outputs. Thus, eXplainable AI (XAI) methods for Human Activity Recognition have emerged to provide intuitive natural language explanations from these models. Different XAI methods generate different explanations, and their effectiveness is typically evaluated through user surveys, that are often challenging in terms of costs and fairness.
%
This paper proposes an automatic evaluation method using Large Language Models (LLMs) to identify, in a pool of candidates, the best XAI approach for non-expert users. Our preliminary results suggest that LLM evaluation aligns with user surveys.


\end{abstract}



\received[Accepted]{for publication at UbiComp / ISWC 2024's XAIforU workshop}

\maketitle

\input{sections/1-introduction}

\input{sections/2-approach}

\input{sections/3-experiments}

\input{sections/4-conclusion}

\begin{acks}
This work was supported in part by MUSA and FAIR under the NRRP MUR program funded by the EU-NGEU. Views and opinions expressed are those of the authors only and do not necessarily reflect those of the European Union or the Italian MUR. Neither the European Union nor the Italian MUR can be held responsible for them.
\end{acks}

\bibliographystyle{ACM-Reference-Format}
\bibliography{references}

\end{document}

%% file: sections/1-introduction.tex
\section{Introduction}

Sensor-based Human Activity Recognition (HAR) is a widely studied research area, and it involves using unobtrusive sensors to identify the activities performed by humans in their daily life~\cite{chen2012sensor}.

The application of sensor-based HAR in smart-home environments is crucial to identify the so-called Activities of Daily Living (ADLs), that are high-level complex activities that humans typically do every day to take care of themselves and maintain their well-being (e.g., cooking, eating, taking medicines, sleeping). Indeed, a smart-home can be equipped with unobtrusive environmental sensors (e.g., motion sensors, magnetic sensors, pressure sensors) to reveal the performed activities. ADLs recognition is crucial in several healthcare applications, including the recognition of abnormal behaviors for the early detection of cognitive decline~\cite{tay2023review}.

The majority of the solutions in this area rely on deep learning methods~\cite{chen2021deep}. Such approaches are effective, but they are often considered as black-boxes mapping input smart-home sensor data into ADLs.
However, non-expert users (e.g., clinicians, caregivers, the monitored subject) need to trust such models by understanding the rationale behind their reasoning.
For this reason, several eXplainable Artificial intelligence (XAI) approaches have been proposed for smart-home ADLs recognition~\cite{arrotta2022dexar, arrotta2022explaining,das2023explainable,jeyakumar2023x}. Such methods, also known as eXplainable Activity Recognition (XAR) systems, are designed to detect human activities and, at the same time, providing clear explanations in natural language for each prediction. The generated explanations indicate which sensor events were considered as important by the classifier to perform a prediction. For instance: \textit{``I predicted that Anna was cooking mainly because she is in the kitchen and the stove is on''}.
However, quantitatively evaluating the effectiveness of the generated explanations is challenging~\cite{mohseni2021multidisciplinary}. Indeed, different XAI methods may lead to completely different explanations, since each approach has a specific mechanism to determine the inner reasoning of the model while making predictions. 

The majority of the works in this area adopted user surveys, by recruiting a large number of subjects to evaluate explanations generated by different approaches. However, this strategy is costly in terms of money, time and human resources. Indeed, recruiting an adequate number of volunteers that guarantees a suitably homogeneous and significant sample, is very hard. A possible solution is to use tools like Amazon Mechanical Turk 
, which enable reaching a much larger audience with the incentive of financial compensation~\cite{das2023explainable,arrotta2022explaining, jeyakumar2023x}. However, besides being costly, such approaches do not guarantee the quality of the results, since the workers do not always provide the necessary attention to the required task. For instance, in~\cite{arrotta2022explaining} the authors introduced some attention questions to exclude bots and users providing random answers just to obtain the reward. Overall, only the $44\%$ of users answered correctly to those questions and could thus be considered reliable. 
Other works proposed metrics to automatically evaluate the quality of the explanations of HAR systems. Specifically, the work in~\cite{arrotta2022dexar} proposed an \textit{explanation score} based on a knowledge-model (i.e., a set of logic rules). Such knowledge-model encodes the relationships between smart-home events and activities, and it is used to evaluate whether the explanation identified the concepts that explain (even partially) the predicted activities. However, this score relies on a rigid model (i.e., an ontology) manually built by domain experts. Building a robust knowledge-model requires significant human effort, and it is questionable whether the resulting model is comprehensive and scalable~\cite{civitarese2019polaris}.

Recently, several research works show that Large Language Models (LLMs) also encode common-sense knowledge about human activities~\cite{ji2024hargpt}. In this paper, we leverage such findings by proposing a novel approach to leverage LLMs for evaluating alternative XAR approaches. Specifically, we propose two different strategies for prompting LLMs in order to evaluate the most effective XAR approach (from a pool of proposals) targeting non-expert users.
By comparing our approach with user surveys on two public datasets, we show that our method ranks XAR methods similarly to non-expert humans.
To sum up, our contributions are the following:
\begin{itemize}
    \item We introduce the novel idea of using LLMs to compare alternative smart-home XAR methods generating natural language explanations, with the goal of selecting the best one.
    \item We propose two different prompting strategies to evaluate explanations.
    \item Our preliminary results suggest that both prompting strategies are aligned with user surveys.
    
\end{itemize}

%% file: sections/2-approach.tex
\section{Evaluating Explanations Using LLMs}


\subsection{Problem Formulation and Research Question}

\subsubsection{Problem Formulation}
Consider a smart-home that is equipped with several environmental sensors (e.g., motion sensors, magnetic sensors, pressure sensors). For the sake of this work, we focus only on single-inhabitant scenarios. The interaction of the user with the home environment leads to the generation of high-level events. For instance, when a magnetic sensor on the fridge fires the value $1$, this implies that the subject opened the fridge.
Let $\mathbf{E}=\{e_1,e_2,\dots,e_n\}$ be the set of possible high-level events captured by the sensors.
The continuous stream of such events is partitioned into fixed-time temporal windows (e.g., using overlapping sliding windows). 
Each non-empty temporal window $w$ includes $k$ seconds of high-level events triggered by the inhabitant. The goal of an eXplainable Activity Recognition (XAR) model is to map each window to the most likely activity performed by the subject, and to an explanation in natural language. Let $\mathbf{A}=\{a_1,a_2,\dots,a_m\}$ be the set of target activities. 
Let $h$ be an XAR model, taking in input a window $w$ and providing as output the most likely activity $a \in \mathbf{A}$ performed by the subject, and a corresponding explanation $exp$. Note that $exp$ describes in natural language the high-level events $E^w$ included in $w$ that are considered the most important to classify $a$ by $h$, and possibly their temporal relationships. In the literature, several examples of such XAR models have been proposed~\cite{arrotta2022dexar,das2023explainable}.

\subsubsection{Research question}
Let $H$ be a set of $K$ different alternative XAR models $h_1, h_2, \dots, h_k$. Let $P$ be a pool of time window. Given a window $w \in P$, all the XAR models in $H$ provide the same output activity but possibly different explanations.
Our research question is the following: can we leverage Large Language Models (LLMs) to choose  the best XAR model in $H$, given the explanations provided to the windows in $P$?

\subsection{Prompting Strategies}
\label{subsec:prompting_stragegies}
In this work, we employed two distinct prompting strategies to evaluate the explanations provided by the different XAR models. Both approaches require providing the LLM with $K$ explanations generated by the $K$ alternative models when processing a window $w$. Sometimes two or more models may provide identical explanations. In such cases, this explanation is provided only once to the LLM, thus providing less than $K$ options. The pseudo-code of the general approach for LLM-based evaluation is presented in Algorithm~\ref{alg:approach}.

%
%

\begin{algorithm}
\tiny
\caption{LLM-based evaluation}
\label{alg:approach}
\begin{algorithmic}[1]
\renewcommand{\algorithmicrequire}{\textbf{Input:}}
\renewcommand{\algorithmicensure}{\textbf{Output:}} 
\REQUIRE 
\quad \\
$P = \{w_0, w_1, ..., w_n\}$, pool of windows
\\
$ H = \{h_0, h_1, ..., h_k\}$, candidate XAR models
\ENSURE 
The best model in $H$ to explain the windows in $P$
\\
\STATE scores = \{$h$: 0 \textbf{for all} $h \in H$ \}

\FORALL {$w \in P$}

\STATE predictions = \{$h$: [ ] \textbf{for all} $h \in H$ \}
\STATE explanations = \{$h$: [ ] \textbf{for all} $h \in H$ \}

    \FORALL {$h \in H$}
        \STATE predictions[$h$].append($h$'s prediction for $w$)
        \STATE explanations[$h$].append($h$'s explanation)
    \ENDFOR
    \STATE $\Vec{S}$ = $\textit{promptStrategy}$(predictions, explanations) 
    \textit{//returns a vector with a score for each model} \\
    \FORALL {$h \in H$}
        \STATE scores[$h$] += $\Vec{S}$[$h$]
    \ENDFOR

\ENDFOR

\STATE \textbf{return} $\argmax_h $ scores[$h$] 
\end{algorithmic}
\end{algorithm}

Given a pool of windows $W$ (e.g., windows collected in a specific time frame in the smart-home), the overall goal is to select the better XAR model. The difference between the two strategies lies in the specific way we asked the model to evaluate them.

\subsubsection{Best-Among-K Strategy}

The first strategy is called the "Best-Among-K Strategy". For each window $w$ in the pool and a set of $K$ alternative models $H$, the LLM is prompted to determine which explanation of $w$ is the best among the ones in $H$. 
Then, we assign a score with value $1$ to the model that provided the best explanation, while the others are scored $0$. When the explanation selected by the LLM is actually generated by more than one model in $H$, we assign a score of value $1$ to all models that generated that explanation.
At the end of this process, the model in $H$ with the highest score is considered the best.

\subsubsection{Scoring Strategy}

In the "Scoring Strategy", we ask the LLM to assign a score to each explanation for a window $w$ using the Likert scale~\cite{joshi2015likert} (i.e., a score between 1 and 5). 
When the same explanation is generated by more than one model in $H$, we assign the same score to all the models that generated that explanation.
At the end of this process, the model in $H$ with the highest score is considered as the best.

\subsubsection{The prompt}

\begin{figure}[h!]
    \centering
    \includegraphics[width=0.6\columnwidth]{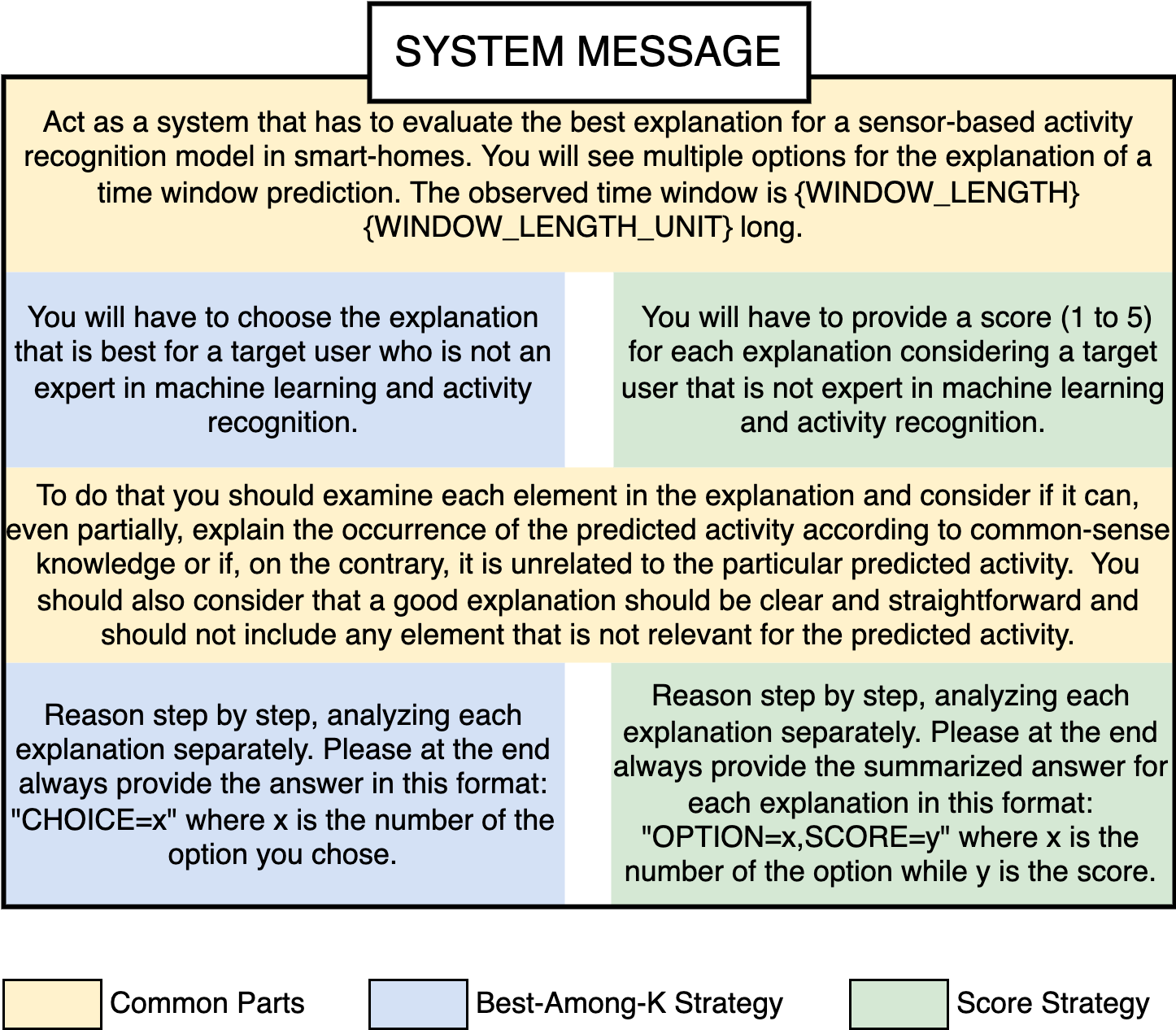}
    \caption{Our System Message}
    \label{fig:system_message}
\end{figure}

\begin{figure}[h!]
    \centering
    \includegraphics[width=0.63\columnwidth]{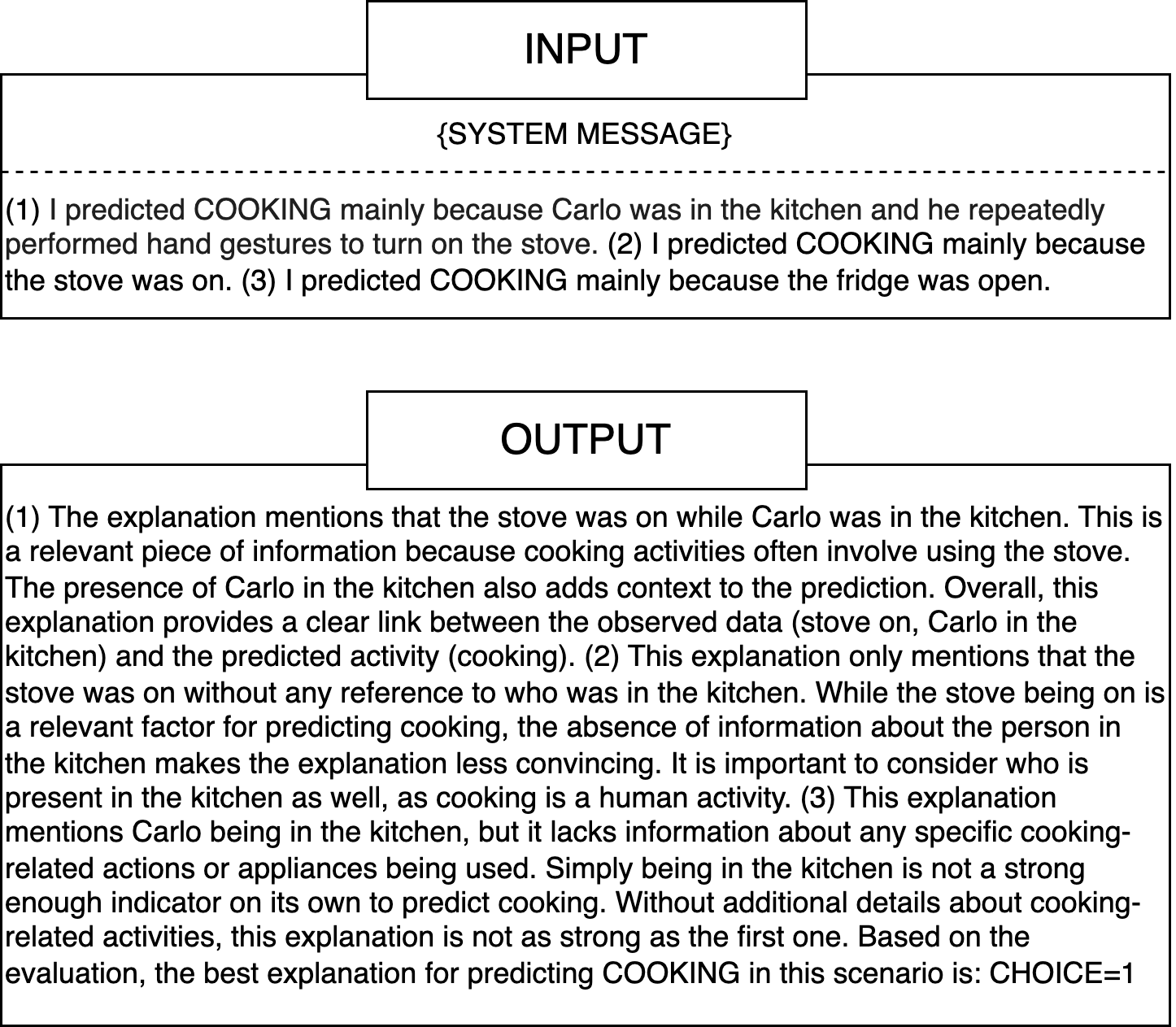}
    \caption{Example of input and output (Best-Among-K)}
    \label{fig:output}
\end{figure}

Figure~\ref{fig:system_message} shows the system prompts (i.e., the instructions for the LLM to perform the task) of our approach. 
For the sake of compactness, the picture depicts the system message for both strategies. Note that the yellow parts appear in the system prompt of both strategies, while the blue parts are only used for the "Best-Among-K Strategy" and the green parts are only used for the "Scoring Strategy". 
Note that the our prompt is the result of a significant prompt engineering effort.  The length of the time window is necessary to provide temporal context, since one window is usually shorter than the typical duration of an activity. We also include instructions to align the LLM-based with the one that a non-expert user would provide. Moreover, we include evaluation criterion inspired by a knowledge-based metric previously proposed in~\cite{arrotta2022dexar}.
The prompt concludes by asking the model to reason step by step, forcing the LLM to adopt the well-known Chain-Of-Thought (COT) prompting strategy \cite{wei2022chain}. 
An example of output of an LLM is depicted in Figure \ref{fig:output}, together with the input that generated it.

%% file: sections/3-experiments.tex
\section{Experimental Evaluation}

In the following, we will introduce the datasets that we considered for our work, the experimental setup adopted, the metrics and, finally, the main results.

\subsection{Datasets}
In order to compare our LLM-based method with user surveys, we obtained from the authors of~\cite{arrotta2022dexar} the surveys that they proposed to real non-expert users to compare the quality of explanations produced by three different XAI methods. Specifically, in their work they considered GradCAM~\cite{selvaraju2017grad} (a saliency based approach), LIME~\cite{ribeiro2016should} (a posthoc explanations approach), and Model Prototypes~\cite{rudin2019stop} (a neural network model designed to be interpretable). 

The authors considered two publicly available Smart-Home HAR datasets to create their surveys: MARBLE~\cite{arrotta2021marble} and CASAS Milan~\cite{cook2009assessing}. The survey prepared on MARBLE considered a pool of 34 windows, while the one for CASAS Milan considered a pool of 27 windows.
The pools were designed to ensure the same number of windows for each activity in the datasets.
In the survey, each window in the pool was represented by the classified activity and an explanation for each XAI approach.
Overall, $84$ users were involved in the MARBLE survey and $63$ for the CASAS Milan survey. We used the same explanations in the survey to evaluate our LLM-based approach and compare it with the user surveys results reported in~\cite{arrotta2022dexar}.
On both datasets, the surveys showed that Model Prototypes generated the most appreciated explanations, while the least appreciated was GradCAM. This result was confirmed in another paper~\cite{arrotta2022explaining}.

\subsection{Experimental Setup}
We implemented our LLM-based system in Python. We experimented with two different models: gpt-3.5-turbo-0125 and gpt-4-turbo, both accessed through the OpenAI APIs. 
The rationale behind the choice of these two models is to investigate whether the results would be consistent between an advanced model like GPT-4 and a cost-effective model like GPT-3.5. We set the models temperature to $0$ to minimize the variability in the answers. 
To construct the prompts and interact with the APIs, we utilized the LangChain library. 
The code is publicly available~\footnote{https://github.com/micheleFiori/llm-xai.git}. Computations were executed in a Google Colab environment. Each experiment was repeated $5$ times to ensure  statistically robust results.

\begin{figure*}[htbp]
  \centering
  \subfloat[MARBLE: Best-Among-K]{
    \includegraphics[width=0.24\textwidth]{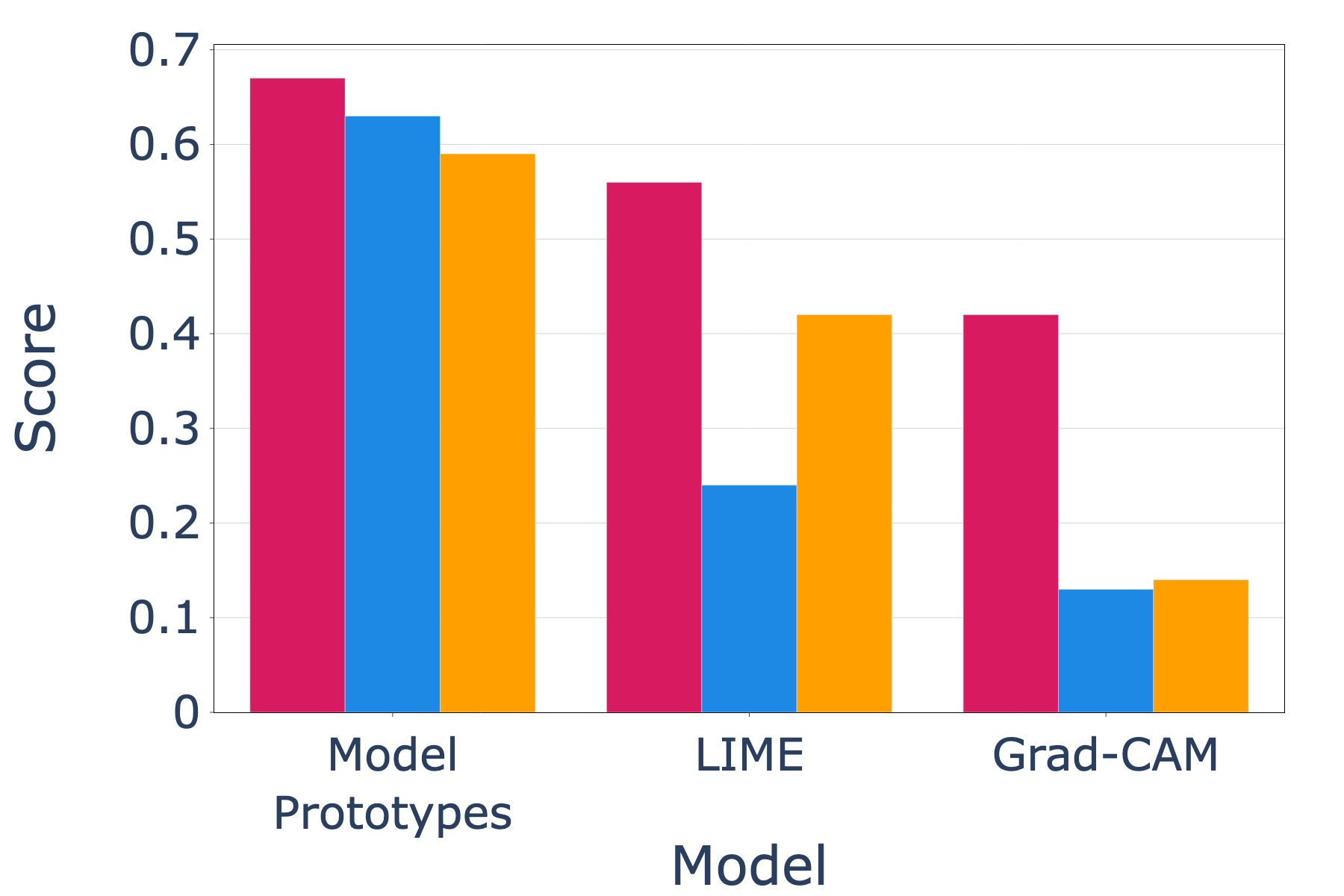}
    \label{fig:marble_best}
}
  \subfloat[MARBLE: Scoring Strategy]{
    \includegraphics[width=0.24\textwidth]{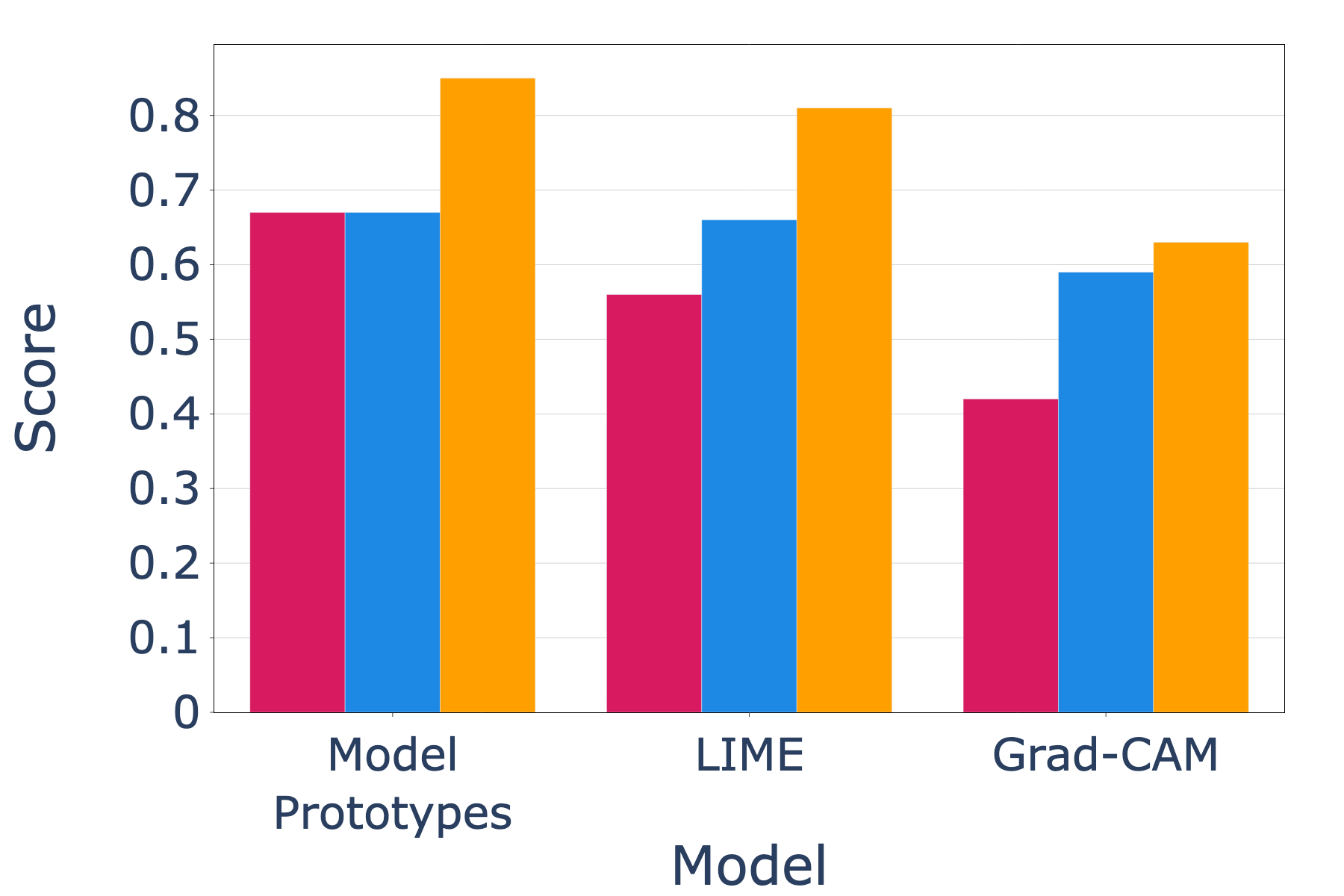}
    \label{fig:marble_score}
  }
     \subfloat[CASAS Milan: Best-Among-K Strategy]{
    \includegraphics[width=0.24\textwidth]{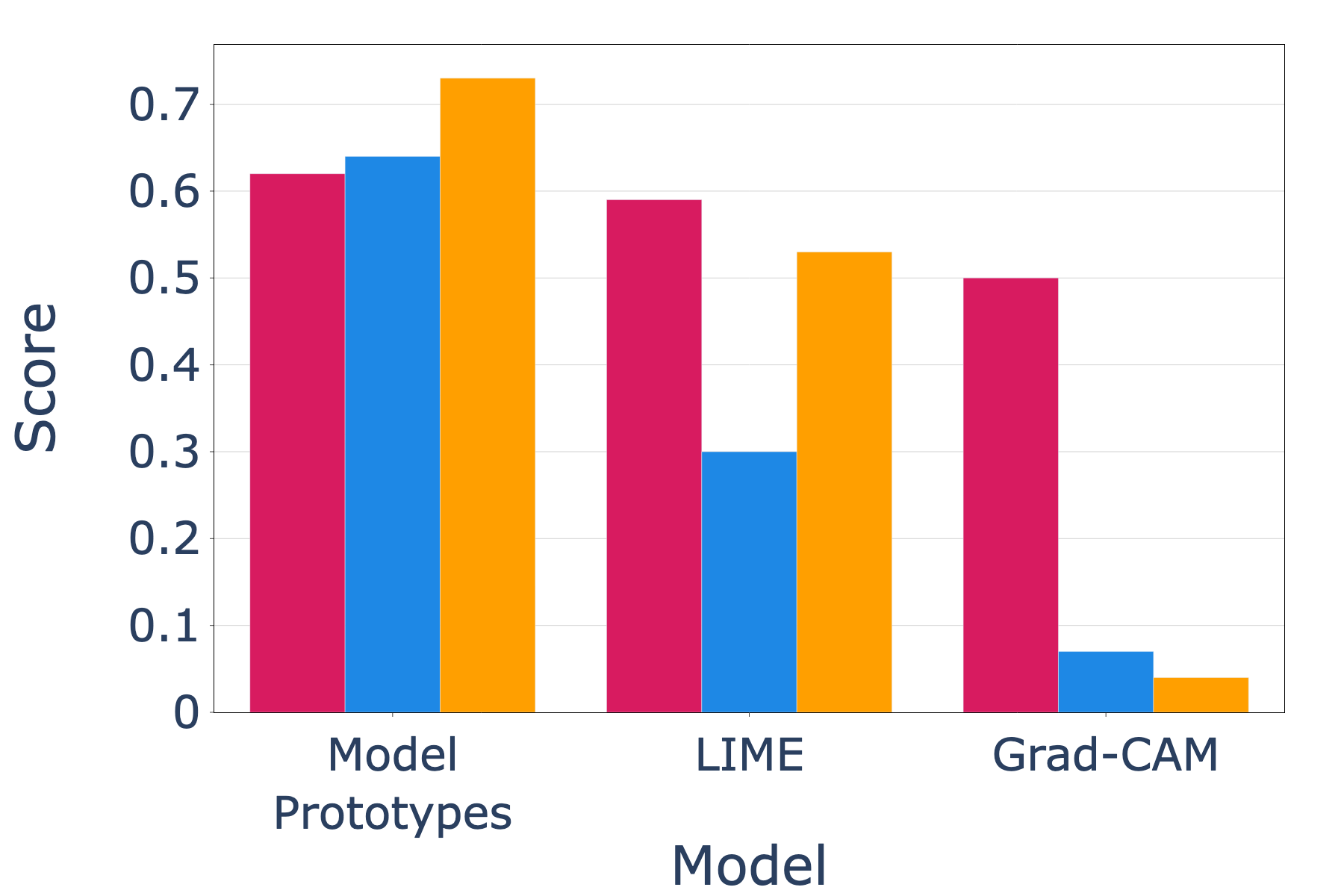}
    \label{fig:casas_best}
  }
  \subfloat[CASAS Milan: Scoring Strategy]{
    \includegraphics[width=0.24\textwidth]{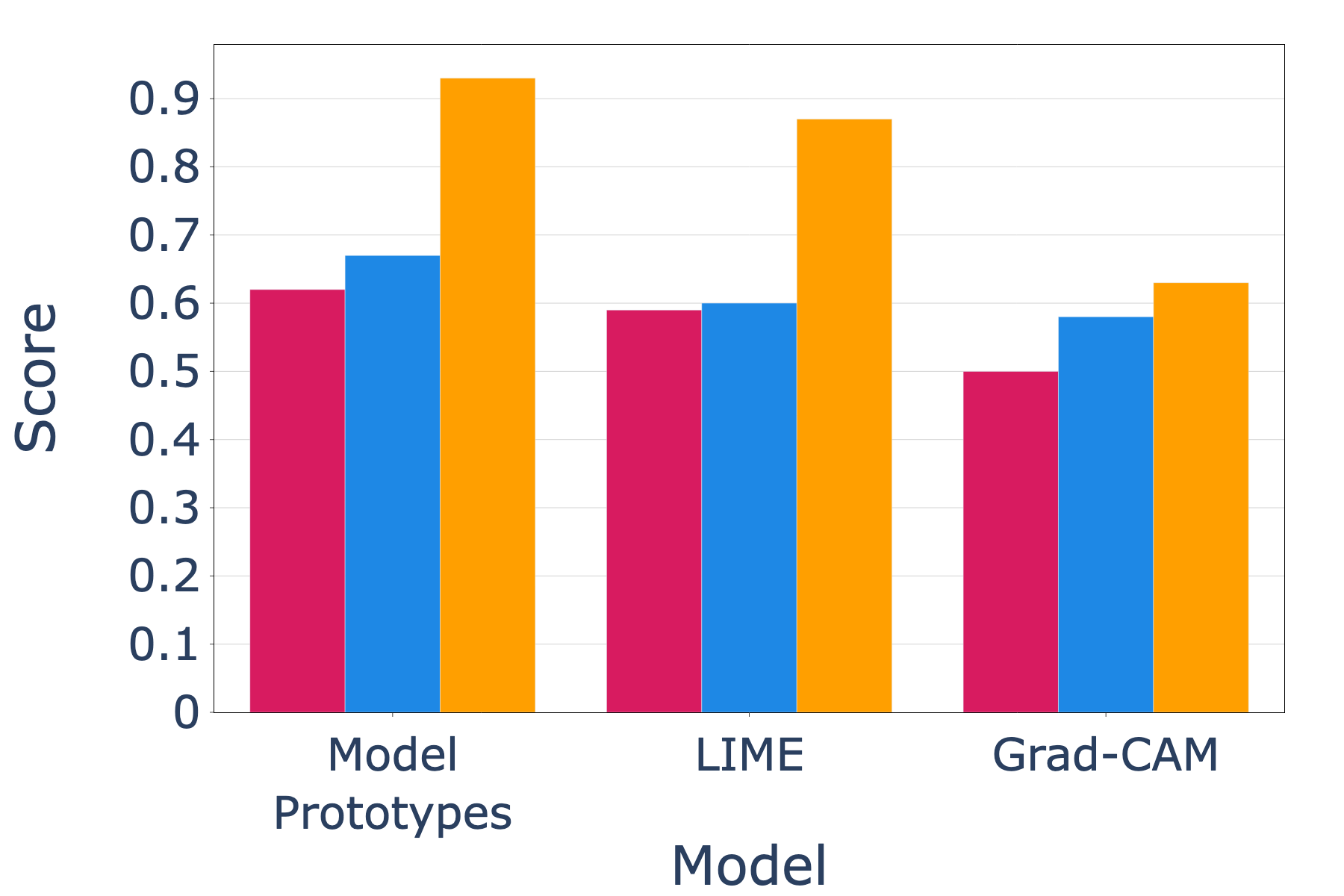}
    \label{fig:casas_score}
  }
  \\
  \subfloat{
    \includegraphics[width=0.2\textwidth]{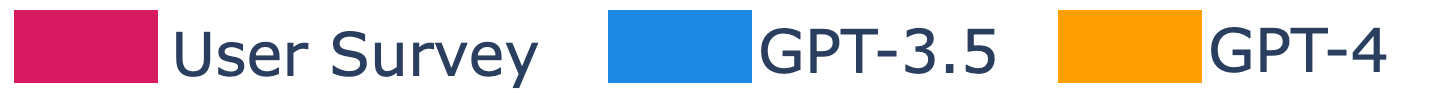}
    \label{fig:casas_score}
  }
  \caption{User survey vs LLM-based evaluation}
  \label{fig:casas_comparison}
\end{figure*}

\subsection{Metrics}
The survey that we use in this paper asked the participants to rate each alternative explanation using a classic Likert scale,  with a grade from 1 (absolutely not satisfying) to 5 (completely satisfying)~\cite{arrotta2022dexar}. The overall result was summarized by computing the average of the scores given by the users for each method and normalizing the results in the interval $[0,1]$. To properly compare our results with the user surveys in~\cite{arrotta2022dexar}, we also normalize the scores in the same range. 

\subsection{Results}

Figures~\ref{fig:marble_best} and~\ref{fig:marble_score} show the results of our evaluation method on the MARBLE dataset considering both prompting strategies. We observe that our LLM-based evaluation is strongly aligned with the one of the user surveys, considering how methods are ranked. Specifically, both user surveys and LLM-based evaluation agree that the best approach is Model Prototypes, followed in order by LIME and GradCam. Note that we are not interested in the absolute values of the scores, but in their relative distance among different XAI methods.
The results also indicate that, among the two LLM models, GPT-4 provides results (in terms of relative proportions between the scores) more aligned with the survey. However, the Best-Among-K strategy tends to penalize GradCam more than the user surveys. This is likely due to the fact that this strategy always assign $0$ points to the ``worst'' explanations. Similar results can be observed in Figures~\ref{fig:casas_best} and~\ref{fig:casas_score} for the CASAS Milan dataset.

Among the two prompting strategies, the one that more accurately reflects the trend of the user surveys considered in the experiments is the "Best Explanation Strategy", although it does not reflect how users rated explanations in the survey. We argue that this is because LLMs are trained using natural language and they are known for struggling when reasoning with numbers \cite{schwartz2024numerologic}.
Finally, while the most advanced model we adopted (i.e., GPT-4) is the one leading to the most similar results compared to the ones in the user surveys, even cheaper models (i.e., GPT-3.5) show a trend in line with user surveys.
The only exception is the Scoring Strategy on the MARBLE dataset, where the score obtained by GPT3.5 on Model Prototypes is only $1\%$ higher than the one obtained on LIME. These results suggests that, in order to reduce costs, even simpler and cheaper model can approximate user surveys.

%% file: sections/4-conclusion.tex
\section{Conclusion and future work}
In this work, we introduced the novel idea of using LLMs to automatically evaluate natural language explanations of XAI methods for sensor-based HAR. Our preliminary experiments show that our prompting strategies make it possible to leverage LLMs to obtain a quantitative assessment that is consistent with more challenging user surveys. 

This is still a preliminary investigation, and we have several plans for future work. First, we will design prompt strategies for different target users. Indeed, in this work we focused on non-expert end-users. However, a similar type of evaluation can be carried out for expert users like technicians or data analysts having a different background. Such expert profiles may need explanations to improve the underlying model or the sensing setup. Moreover, since in this paper we only evaluated whether the explanation is in line with common-sense knowledge about the predicted activity, we will investigate many other aspects that are crucial for explanations (e.g., understandability, trustworthiness, reliability)\cite{mohseni2021multidisciplinary}.